\documentclass{PoS}

\title{The VLBI morphology of M\,81$^*$ at 43\,GHz}

\ShortTitle{$\lambda$7\,mm VLBI Morphology of M\,81$^*$}

\author{{Eduardo Ros}\\
        Max-Planck-Institut f\"ur Radioastronomie, Auf dem H\"ugel 69, D-53121 Bonn, Germany\\
        E-mail: \email{ros@mpifr.de}}

\author{Miguel A. P\'erez-Torres\\
        Instituto de Astrof\'{\i}sica de Andaluc\'{\i}a-CSIC, Camino Bajo de Hu\'etor 50, E-18008 Granada, Spain \\
        E-mail: \email{torres@iaa.es}}

\abstract{
The radio source M\,81$^*$ at the core of the nearby spiral galaxy
M\,81 is a low-luminosity active galactic nucleus.
The close distance of 3.63\,Mpc allows its morphology to be studied in
great detail.  Here we present preliminary results from 
continuum $\lambda$7\,mm 
VLBI observations of its core, using phase-referencing techniques.  
These observations set constrains on the size of M\,81* at this 
frequency and enable us to test the frequency dependence on its
physical properties.
}

\FullConference{The 9th European VLBI Network Symposium on The role of VLBI in the Golden Age for Radio Astronomy and EVN Users Meeting\\
		 September 23-26, 2008\\
		 Bologna, Italy}

\begin{document}

\section{Introduction}
At the heart of the nearby galaxy 
M\,81 (B0951+693, J0955+6903, NGC\,3031; $D$\,=\,3.63\,Mpc),
the low-luminosity (10$^{37.5}$\,erg\,s$^{-1}$)
radio source M\,81$^*$
is one of the closest active galactic nuclei (AGN).  
This object is four orders of
magnitude brighter than Sgr\,A$^*$, but has 
similar shape and polarization properties.
M\,81$^*$ shows a stationary feature in its
structure with a jet to the northeast (seen at 
$\lambda$3.6\,cm; \cite{markoff08}).
Its size follows the law $\theta\propto\nu^{-0.8}$ between
2.3\,GHz and 22\,GHz.
It shows also a frequency-dependent orientation with 
its position angle (P.A) varying from $\sim$40$^\circ$ at 22\,GHz to
$\sim$75$^\circ$ at 2.3 GHz \cite{bietenholz00}.
M\,81$^*$ is variable at all wavebands, and presents 
relativistic X-ray iron lines at X-rays \cite{markoff08}.

\section{Observations and data analysis}
\begin{figure}[t!]
\centering
\includegraphics[bb= 52 173 589 675, clip,width=.75\textwidth]{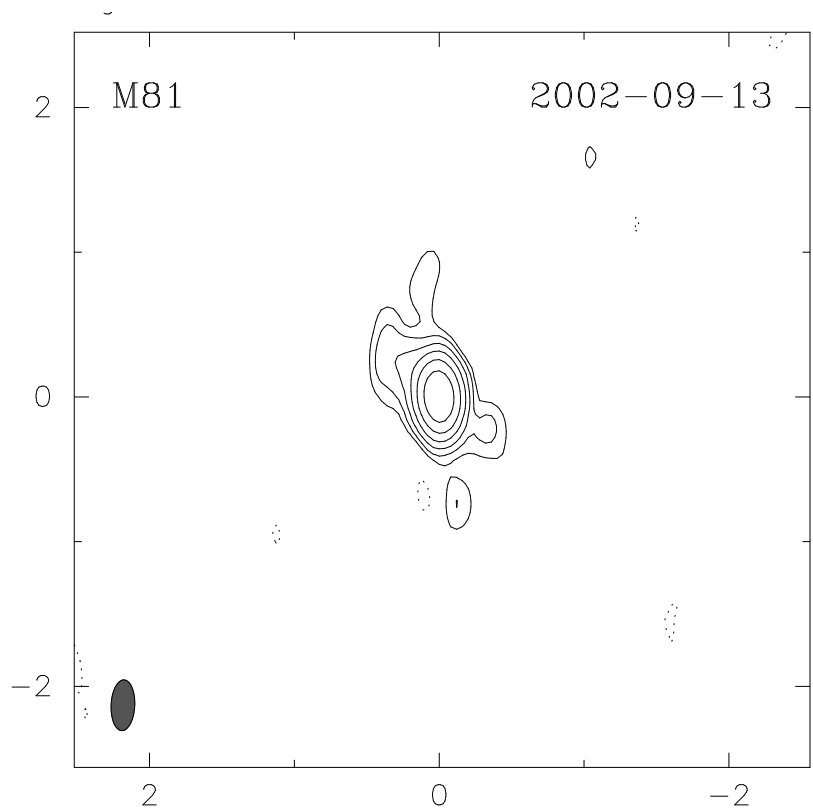}
\caption{
Hybrid contour map of M\,81$^*$ from the
self-calibrated data set (see Text). The lowest
contour is of 1.57\,mJy\,beam$^{-1}$. The
peak of brightness is of 98.5\,mJy\,beam$^{-1}$. 
The beam size, shown at
the bottom, left, is of 0.351$\times$0.164
mas, in P.A.\ 2.6$^\circ$.
}
\label{fig1}
\end{figure}

We observed M\,81$^*$ on September 13, 2002
with the complete Very Long Baseline Array at a
frequency of 43\,GHz ($\lambda$7\,mm).  Observations
were performed in phase-referencing mode, using the 
BL\,Lac object B0954+658 (J0958+6533; $z$=0.368), 
34\,arcmin apart on the sky.  After calibration,
the first \textsc{fring} run (NL was used as reference
antenna) on the target source
provided a 65\% failure in the overall solutions, while
an 80\% of successful fringe detection were obtained
for the reference source.
Hybrid mapping was performed for the calibrator source
using \textsc{difmap}.  The resulting self-calibration
solutions for the amplitude were applied to the M\,81$^*$
data back in \textsc{aips}.  A run of \textsc{imagr}
in \textsc{aips} was performed, yielding an image.  
The \textsc{clean} table of those was used as input
in \textsc{fring} to make structure-free delay and rate
solutions.   Using these values we restricted the
search windows for delay and rate in M\,81$^*$, and we
applied \textsc{fring} again, providing very satisfactory
results.  We then were able to perform a hybrid map
of M\,81*, shown in Figure~\ref{fig1}.
Our image
provides hints of extended emission to the north-east, in
agreement with \cite{markoff08}.

\section{Discussion}
We also model fitted the visibility data of M\,81$^*$
with a Gaussian elliptical function.  
The source has an elliptical shape extended in the 
north-east--south-west direction.  The major axis of the function is
of 93$\pm$1\,$\mu$as, with an axis ratio of 0.3$\pm$0.1.
We used the model fitted size of our target source to compare it with 
earlier measurements.
A power-law fit to the sizes as a function or frequency, 
(including the epoch-averaged
measurements by \cite{bartel82,bietenholz96})
together with our
measurement
shows a
dependence of $\theta\propto \nu^{-0.88\pm0.05}$.
These
results are compatible with the
ones previously reported by \cite{bietenholz00}.

Our observations have resulted in the 
detection of M\,81$^*$ at the shortest
wavelength performed with VLBI so far.  Further
observations are needed to discern the nature of the
jet in M\,81$^*$.  Future monitoring observations could
also trace any structural changes and/or proper motions in the jet.

\begin{small}
\paragraph*{Acknowledgments}
MAPT is funded by a Ram\'on y Cajal contract of the 
Spanish Ministry of Education. The Very Long Baseline
Array is operated by the USA National Radio Astronomy 
Observatory, which is operated by Associated
Universities Inc., in cooperative agrement with the 
USA National Science Foundation.
\end{small}

\end{document}